\documentstyle[prl,aps,twocolumn]{revtex}
\tightenlines
\begin{document}
\twocolumn[\hsize\textwidth\columnwidth\hsize\csname
@twocolumnfalse\endcsname
\draft
\title{Will Quantum Cryptography ever become a successful technology
in the marketplace?}
\author{Hoi-Kwong Lo}
\address{MagiQ Technologies Inc.\\
275 Seventh Avenue\\
26th Floor\\
New York, NY 10001}

\date{\today}
\maketitle
\begin{abstract}
We assess the potential of quantum cryptography as a technology.
We highlight the fact that academia and real world
have rather different perspectives and interests. Then, we
describe the various real life forces (different types of
users, vendors of crypto-systems,
conventional cryptographers, governments) behind the decision of
the adoption
(or rejection) of quantum cryptography and their
different interests.
Various roadblocks to the widespread application of
quantum cryptography
are discussed. Those roadblocks can be fundamental, technological,
psychological, commercial or political and many of them have
nothing to do with the security of quantum key distribution.
We argue that the future success of quantum cryptography as a technology
in the marketplace
lies in our ability to appreciate and to
overcome those roadblocks and to
answer real world criticisms on the subject.

\end{abstract}
\pacs{PACS Numbers:}
]
\narrowtext
\section{Introduction}
\label{Intro}

In this quantum cryptography workshop\footnote{This paper is an
extended version of a talk to be presented in NEC Princeton workshop
on quantum cryptography, Dec. 13-15.} we
have heard many interesting talks. From an academic point of
view, it is quite clear that
quantum cryptography is a
very active research area. Now, from the technological point of view,
it is natural to ask about the potential of quantum cryptography as
a {\it technology}. In other words, will
quantum cryptography ever be widely used in future?

Given the immense progress in both the theoretical and
experimental sides in the last few years,
some of us in the audience may be tempted to say `yes'.
However, ultimately the answer to this question does not depend
on the subjective opinions of research scientists, but on the
complex social and economic forces behind it as well as
future advancements in the quantum technology.

On the question of whether
quantum cryptography will ever be
widely used in future, I certainly
do not claim to have a full answer.
In this talk, I will simply
share with you some of my thoughts on the
subject.
None of the viewpoints expressed here are original. Nor are they
sophisticated. They are just my personal simplifications
and understandings/misunderstandings of what is
well known to people in other walks of life.
However, those viewpoints may not be commonly known to
quantum cryptographers.
Being an industrial researcher with a non-negligible amount of
experience in research and development of real-life conventional
security systems, I find it of value to introduce these viewpoints to
other quantum cryptographers. My hope is to stimulate
further discussions on the subject. Your comments, corrections and
criticisms will be welcome.

\section{Academia vs Real World}
\subsection{Technology focussed vs solution focussed}
Let me begin by saying that the academic world and the real world
have very different perspectives. In academia, we often deal with
curiosity driven research. Even in quantum technology,
our main focus is {\it technology}. That is to say
the technological aspects of a subject.
For instance, in this workshop many talks deal with the
fundamental and technical issues of the security of
quantum cryptographic systems. Important as they
are, those subjects are so esoteric that
they are quite beyond the understanding of
even the most sophisticated developers and
customers of conventional cryptography.
More importantly, those subjects do not necessarily
address their real world concerns.

In the real world, customers (users of cryptography) generally
have problems and they look for {\it solutions}, not technology.
It does not matter whether it is high-tech or low-tech, so long
as it can solve their problem, they will take it.
For instance, putting an eraser on top of a pencil is
a trivial idea from the technological point of view. From the
users' point of view, this can be regarded
as a major invention that offers convenience and
added value to the individual eraser and the pencil.
As another example, nuclear power generators may be a high-tech
solution. But, it must compete with low-tech alternatives
like oil and coal in a competitive commodity market---electricity
generation.

Clearly, customer acceptance is very important to the success of
a technology in the marketplace. Beside         
customers,
there are many other players
in the development/adoption
of new technologies.
Different players
have different interests and concerns, some of which may be
regarded as irrational by outsiders. Whether we like it or not,
the only way to assure that a technology is adopted is to better
understand the diverse interests and concerns of different players
in the field.

\subsection{Players in the Quantum Game}

Let me introduce the interested parties in the development/adoption
of cryptographic/security systems one by one and describe their main
concerns.

{\bf 1. Academia}

(A) {\bf Quantum Cryptographers Type I (Particularly
Theorists)}: The main interest of
theoreticians in quantum cryptography is to design cryptographic
protocols with {\it perfect security} \cite{perfect}.

(B) {\bf Quantum Cryptographers Type II (Particularly
Experimentalists)}: The main
interest of experimental quantum cryptographers is to design and
implement quantum cryptographic schemes that are feasible with
current (or near future) technology and secure against {\it realistic
attacks} \cite{argue}.

(C) {\bf Conventional Cryptographers?} Some people may
argue that many conventional cryptographers live in the same
academic world as quantum cryptographers. I have no comment
on this argument.

{\bf 2. Real World}

(A) {\bf Users Type 1 (individuals)}: The main interest of
individual customers in using cryptographic/security product is
often the {\it peace of mind}. If the users voluntarily use the
product, this peace of mind may be due to the {\it preceived}
security offered by the product. If the users are forced to use
the products by others, the peace of mind may arise because
they make their employers happy.

Cost and transparency are two other major concerns of
the individual customer. Someone working on information security
once told me
that the general feeling in the community is that
security does not sell.
(i.e., While customers worry about security, they are
unwilling to pay for a higher-price
product for the sole reason of its being
more secure.) The acceptable additional cost of a more
secure product is essentially zero.
People are interested in a solution (say a payment scheme)
that is offered as
a complete package: versatility, convenience of use,
reliablility, cost and security. ''Security'' is just a small term
in the whole equation. Here, I have put security in a quotation mark
because it is {\it preceived} security that counts. A layman
generally does not understand real security. Besides, it
appears to me that
there is no logical consistence in users behaviors when
many of them seem perfectly happy in giving out their
credit card numbers over the phone, but not over the Internet.

Transparency of the operation of encryption is also a plus.
While a layman can intuitively appreciate the security offered by a
an encrypted file which looks garbled even to the unsophisticated
eye, the same cannot be said for
quantum cryptography.

(B) {\bf Users Type II (businesses)}:
A notable motivation for
many businesses such as the banking industry to employ cryptographic
products for its customers is to limit its financial and legal liability.
Businesses generally accept a certain degree of
financial losses due to insecure products as parts of their
normal
operating
cost in doing businesses. Therefore, non-perfect security of
conventional cryptographic systems is not
a bad thing, but a fact of life. The important things are to have
risk management and to have risk factors that are well understood.
Employing industrial standards is very useful in reducing businesses'
financial and legal liability. Employing a non-standard disruptive
technology like quantum cryptography is much more risky.

Securing {\it long} distance communication is an important
concern in businesses. As international companies are now getting more
and more global and tremendous amount of data are
passed between different offices of the same company or
different companies, there is an increasing need in securing those massive
transcontinental communications. Besides, there is an increasing need
for {\it post-Cold-War} type of applications like authentication
and signatures.

(C) {\bf Vendors of Crypto Products}: Like any
other businesses, the main concern of
vendors of crypto products is to make money in the long run.
Besides, vendors have vested interests in deciding which
technology to employ. For instance, a vendor with a large number of
patents and
products in the elliptic curve crypto-systems might be tempted to
emphasize the strengths of elliptic curve crypto products
compared to products based on other principles.

(D) {\bf Conventional Cryptographers and Security Experts}:
Because of their own background and experience, conventional
cryptographers and security experts are keen to use something
that they can understand and trust such as the one-way
function hypothesis. If you ask them whether they
believe in quantum mechanics or one-way hypothesis more,
their answer is clear.

(E) {\bf Governments}: Different departments in a government
have different interests. For instance, the military and the
foreign office are certainly interested in having perfect security
for their communications. On the other hand, for agencies such as
the FBI, the ability to wiretap communications of the criminals is
very important. From this point of view, perfect security might
threaten
national security and should be
discouraged or controlled by laws.
I am not up to date with the current US regulations. However,
until recently, cryptography has been regarded as
ammunition in the US laws,
subject to the strictest control in its usage and
export.

\section{Roadblocks}

Having introduced the different players and their interests in
cryptography, it is the time to discuss the major roadblocks to
the future deployment of quantum cryptographic systems.
For ease of discussion, I will divide those roadblocks into
different classes. However, my division is somewhat subjective.

\subsection{Fundamental roadblocks}

Quantum cryptography is a fundamentally limited technology.

(A) {\bf Impossibility of unconditional security for many applications}

First, it has a limited range of applications.
The fundamental appeal of quantum cryptography has been
perfect or unconditional security (i.e., security
guaranteed by the laws of quantum mechanics only and
without making any computational assumptions).
However, the unconditional security of
a number of important basic protocols such as
bit commitment \cite{bit},
one-out-of-two oblivious transfer \cite{secure}
and one-way identification (and more generally,
one-sided two-party secure computations) \cite{secure} have
been shown to be
impossible in a series of no-go theorems.
What it means that all such protocols must
require quantum computational assumptions.

(B) {\bf Lack of public key based quantum cryptographic schemes}

Second, quantum cryptography has made no significant
contribution to
public key cryptography.
Many real life cryptographic applications such as
signature and authentication schemes in the Internet age
involve public key cryptography. However, very little (if anything)
has been done on quantum cryptographic signature and authentication
schemes that are public-key based.

\subsection{Technological roadblocks}

(C) {\bf Limited distance in current quantum key distribution experiments}

Experimental quantum key distribution has been performed over
tens of kilometers. However, a major market for secure
communication is, in fact, transcontinental communications.
Until the distance achieved in experimental quantum key
distribution increases by two order of magnitude, quantum
key distribution is not a feasible technology for this major
market sector.

(D) {\bf Limited data rate}

The current data rate for experimental quantum key distribution is
of the order kbits for second. (Worse still, the post-processing
including error correction and privacy amplification is quite
massive.) In contrast, the current
world record for a {\it  single mode} optical fiber communication is
160 Gbits for second \cite{record}.
``Multiplying 160 gigabits over additional
wavelengths, we expect to be able to scale up to many trillions of
bits a second in the foreseeable future.'' says Alastair Glass,
director of Bell Labs Photonics Research Labs.
If quantum key distribution is ever going to be widely
used for one-time pad
application for the massive data being transmitted in commercial optical
fibers, there is probably a ten order of magnitude gap in data rate to
be closed in the foreseeable future.

\subsection{Commercial roadblocks}

(E) {\bf Equipment size is too big.}

Ideally, cryptographic applications should be done either by a
software or a very small hardware component such as a smart card
or a CD. Unfortunately, current quantum cryptographic systems are
quite big. Shrinking a quantum cryptographic system to the size of
a briefcase is already a big challenge. Shrinking it to the
size of a smart card requires much ingenuity and development.

(F) {\bf Cost is too high.}

The acceptable additional cost of a more secure cryptographic
product for individual consumers is essentially zero
while the components of existing quantum cryptographic system
cost hundreds or even thousands of dollars.

(G) {\bf Integration with existing infrastructure in information
technology requires further developments.}

Except for niche markets, we cannot expect an optical fiber to
be solely dedicated to quantum communications for any substantial
period of time.
The integration of quantum technology with
conventional and existing infrastructure in information technology
requires much further work.

\subsection{Security roadblocks}

(H) {\bf Known loopholes in current implementations}

While quantum cryptography claims to offer perfect security in
theory, in practice current experimental implementations contain
quite a number of security loopholes. It has been argued that
essentially none of the existing implementations is actually secure
\cite{argue}.
Plugging those known loopholes is a highly non-trivial experimental
and theoretical design problem.

(I) {\bf Hidden loopholes in implementations}

All security analyses of quantum cryptographic systems involve
idealizations. 
It is
highly probably that many other fatal security
loopholes in the implementations of quantum cryptography remain to
be discovered. Given the slippery nature of the
subject, quantum cryptography hardly inspires the confidence of
potential users.

The best way to construct a secure cryptographic system is to
try hard to break it. Unfortunately, until recently very few
people worked on
breaking
quantum cryptographic systems. Without an army of people trying to
break them, the security of quantum cryptographic systems are largely
untested.

\subsection{Psychological/
Vested Interest roadblocks}

(J) {\bf Conventional cryptographers have no confidence in
quantum mechanics.}

Most conventional cryptographers do not understand quantum
mechanics. Nor are they familiar with its many applications.
In any case, the burden of proof of the usefulness of
a new technology lies on its own practitioners, not conventional
cryptographers.    In contrast, from their point of view,
things like the one-way function hypothesis, the hardness of
factoring are well-tested principles and something that they
understand well.  It is wishful thinking to ask them to
take a leap of faith by abandoning their well cherished
philosophy and taking up a black box philosophy for
no apparent good reason.

Moreover, Neal Koblitz remarked in
Crypto' 97 (the most important international conference in
fundamental research in cryptography)
that many
cryptographers hate quantum computation
because if it flies,
it will put many of them out of business.
Indeed, if a quantum
computer is ever built, many public key cryptographic
schemes that are widely used today will be totally unsafe.
This could potentially kill public key cryptography and
throw cryptography back to the ``dark age'' \cite{smart}---a nightmare
scenario for electronic commerce and data security. 
[See, however, \cite{brassard}
for a discussion of the possibility that public key
cryptography may actually survive quantum attacks.]
Since quantum cryptography is a part of quantum information
processing, it is only natural that conventional
cryptographers may not like it neither.

(K) {\bf Vendors of conventional cryptographic products
have vested interests
in promoting and preserving conventional technologies.}

If this is what conventional cryptographers might think
as individual researchers, you can imagine what existing
crypto-system vendors might think about quantum cryptography:
Quantum cryptography is far more likely to be seen as an unwelcome
threat rather than a potential opportunity.

In the history of technological developments, disruptive
technologies are often made possible by new firms rather than
existing
firms that have large stakes in the dominant
existing technology.

\subsection{Political/Legal roadblocks}

(L) {\bf Quantum cryptography may be limited by
governmental
crypto control policies.}

As mentioned earlier, in the US, usage/export
of strong
cryptography is subject to
stringent governmental
control. Any future usage/export of quantum
cryptographic systems will be subject to the same
stringent set of regulations. How to reconcile the
main selling point of quantum cryptography (strong
security) and cryptography control (limitation on the
employment of strong cryptography) is a subject that
deserves future investigations.

\section{Future Directions}

Having discussed the various roadblocks to the future
widespread applications of quantum cryptography, I hope that
you will agree that the issue of commercial feasibility
is much more complicated than
a research scientist may naively think.
Certainly, my own grasp of the problem is limited.
If there is a lesson in this talk, it is the following:
To better understand the issue of commercial feasibility,
it is best for quantum cryptographers to engage more
in constructive conversations with people in the real world
(users, vendors, conventional cryptographers, government officials, etc).
While we
do not have to agree with what they say, it is important for us
to understand their views clearly. The future adoption of
quantum cryptography relies on their acceptance.

On the more technical side, I offer the following subjective list of
future directions.

{\bf 1. Develop new applications for quantum cryptography}:

In my opinion, it is important to develop new applications of
quantum cryptography such as signature and authentication schemes,
quantum voting, etc. Since various no-go theorems have ruled out
the possibility of a number of cryptographic primitives, future
quantum cryptographic systems
may well be based on quantum computational assumptions \cite{bennett}.
Therefore, it
would be of practical interest to invent a ''quantum one-way trapdoor
function''and public
key based quantum cryptographic systems.

One possible viewpoint to take is to regard quantum cryptography
as a natural extension (rather than a replacement) of conventional
cryptrography and put its foundation on computational assumptions on
both conventional cryptography and quantum mechanics.
How to combine the advantages offered by quantum mechanics and
public key infrastructure is a big issue. It would be of particular
interest to construct a public-key based
quantum encryption scheme and show rigorously
that breaking it will require the {\it
 simultaneous} breaking of widely accepted
assumptions in both conventional cryptography (such as cracking
the Diffe-Hellman key exchange scheme)
and quantum computation (such as the ability to achieve quantum
computation/measurement
involving more than $N$ qubits).
Such an encryption scheme will convince people in both
conventional and quantum cryptographic communities
that it is
secure against any realistic attacks.

[Brassard \cite{brassard} has emphasized the possibility that conventional
public key cryptography may actually survive quantum attacks. According to
\cite{brassard}, it has been argued in \cite{BBBV} that quantum resistant
one-way function that can be computed efficiently with {\it classical}
computers but cannot be inverted efficiently even with a quantum
computer may well exist. That would be bad news for quantum
cryptographers,
though.]

{\bf 2. Use teleportation \cite{tele} to plug
security loopholes \cite{LC99}}:

A major criticism on quantum cryptography is that it
may contain many hidden security loopholes.
For instance, while it is often assumed that a
photon source emits single photons, in real life perfect
single photon sources are notoriously hard to make. Besides, experimental
systems generally contain higher energy levels whose
occupancy is totally ignored in most security analysis.
Indeed, as emphasized by,
for example, John Smolin \cite{smolin}, it is
even conceivable in principle that
an eavesdropper can hide a quantum robot in the
quantum signals received by the two
users. Owing to this quantum Trojan Horse problem,
quantum cryptographic systems seem inherently
unsafe.

Nonetheless, one can argue that
by using teleportation,
quantum cryptographic systems can be made no more
unsafe than conventional ones \cite{LC99}.
One can reduce the quantum Trojan Horse problem to a
conventional Trojan Horse problem. This is done by the
following method.
Instead of receiving any untrusted quantum signals from a
quantum channel, each user insists that any signal
should be teleported to him/her. For instance, Bob
prepares locally
an EPR pair and sends a member to a laboratory outside his
door. Any incoming quantum signal will be teleported to him
by his doorman outside his door. What he receives are just
classical messages. Note that teleportation provides an
exact counting of the number of
dimensions of Hilbert space of the
reconstructed state. This is so even if the original
EPR pair that Bob prepares is imperfect and contains hidden
dimensions.

Of course, the problem of classical Trojan Horse attack remains.
But, this is inevitable. Since Bob's goal is to receive
classical communications from Alice through an untrusted channel,
if receiving untrusted classical
messages
is a problem, the whole enterprise of secure communication is
simply hopeless.

{\bf 3. Use quantum repeaters \cite{repeaters} to
extend the range of secure quantum
key distribution.}

This is crucial if quantum key distribution is ever to make
any impact on intercontinental communication.

{\bf 4. Increase data rate for quantum key distribution.}

Existing schemes for quantum key distribution such as BB84 and Ekert's
scheme
are based on
two-level quantum systems and as such their data rates are limited.
If quantum cryptography is ever widely used as one-time pad for
encrypting massive data in communications, higher level systems
and particularly continuous variable quantum cryptography are a
way to go forward. This would mean
that many of the current investigations may become obsolete in the
near future.

{\bf 5. Miniaturization}.

The ultimate goal is to reduce the size of quantum
cryptographic systems to that of a smart card or a 
compact disc.

{\bf 6. Integration with existing infrastructure in information technology.}

It may be hard to justify the cost of construction of an entirely
new infrastructure dedicated to the long-distance transmission of
quantum signals. Integration of quantum technology with existing
infrastructure (including optical fibers) in information technology
is, therefore, an important subject.

{\bf 7. Towards an international standard for quantum cryptography.}

Ultimately, some form of international standards will be needed for
the widespread deployment of quantum cryptography.

{\bf 8. We need quantum hackers.}

We have seen encouraging signs that researchers are finally
taking a critical look at the security of
current experimental
implementations of quantum cryptographic systems \cite{argue}.
In order to better understand the real risk of employing
quantum cryptography, much more
should be done on the subject.

An attacker should attack a Chinese Wall
from its weakest point.
The weakest point of a cryptographic system often lies in the
blindspot of its designers. A cryptographer may regard
a cryptographic system as a mathematical
black box function which
provides an output for each input. However, in real life
the box is never black to begin with.
(Private keys embedded in a smart card circuitry may be
read out by illuminating the smart card with various
wavelengths of electromagnetic radiations.)
The black box also gives out timing information, power
consumption information, etc, etc.
The inputs to the black box include also
its power supply, something that
is subject to manipulations by
malicious parties. The designer of the black box may try to
cheat by designing a black box that leaks information in
a subtle encrypted way that can be read by only the designer.

Indeed, in
conventional cryptography, it is often the case that
the most powerful attacks against a system has little to do
with the fundamental design or mathematical equations underlying
the design. The devil is in the actual
implementation, rather than
the fundamental design. If we are really interested in the
future of quantum technology, we must face up to
those subtle
loopholes in implementations. A way to do so is to become a quantum
hacker and devise innovative methods of cracking experimental
quantum cryptographic system.

{\bf 9. Crypto control of quantum cryptography?}

The issue of cryptography control of quantm cryptography remains to
be addressed. I have no particular suggestion.

\section{Acknowledgment}
I have greatly benefitted from helpful discussions with many
colleagues,
collaborators, and experts in both conventional cryptography/security
systems
and quantum cryptography. I would like to thank them all and
apologize for any misrepresentations of their ideas/observations
in this talk.


\begin{references}
\bibitem{perfect} See, for example, H.-K. Lo and H. F. Chau,
Science 283, 2050 (1999) and supplementary material at
www.sciencemag.org/feature/data/984035.shl
and D. Mayers,
Los Alamos preprint archive (LANL) quant-ph/9802025 version 4.
\bibitem{argue} See, for example,
G. Brassard {\it et al.}, Los Alamos preprint archive
quant-ph/9911054.
\bibitem{bit} D. Mayers, Phys. Rev. Lett. {\bf 78}, 3414 (1997)
[also in Los Alamos preprint archive (LANL) quant-ph/9605044];
H.-K. Lo and H. F. Chau, Phys. Rev. Lett. {\bf 78}, 3410 (1997)
[also in LANL quant-ph/9603004].
For a review, see, for example,
H. F. Chau and H.-K. Lo, Fortsch. Phys. {\bf 46},
507 (1998) [also in LANL
quant-ph/9709053].
\bibitem{secure} H.-K. Lo, Phys. Rev. {\bf A56}, 1154 (1997).
\bibitem{record}
http://www.bell-labs.com/news/1999/november/10/3.html
\bibitem{smart} Such a viewpoint has been relayed to me
by Nigel Smart, private communications.
\bibitem{brassard} G. Brassard, in {\it Advances in Cryptology---Crypto' 97},
Springer-Verlag, LNCS 1294, p. 337 (1999).
\bibitem{bennett} C. H. Bennett, private communications.

\bibitem{BBBV} C. H. Bennett, E. Bernstein, G. Brassard and U. Vazirani,
{\it SIAM Journal on Computing}, ?. 
\bibitem{tele} C. H. Bennett {\it et al.}, Phys. Rev. Lett. {\bf 70},
1895 (1993).
\bibitem{LC99} See H.-K. Lo and H. F. Chau, Note~21
of Ref.~\cite{perfect}.
\bibitem{smolin} J. Smolin, private communications.
\bibitem{repeaters} See, for example, H.-J. Briegel, W. D\"{u}r, S. J.
van Enk , J. I. Cirac and P. Zoller, Philos. Trans. R. Soc. London Series
{\bf A356}, 1713 (1998).

\end{references}
\end{document}